\documentclass[doublecol,figures]{epl2} 

\usepackage{amsbsy,amssymb,amsmath}
\usepackage{epsf,colordvi,graphicx,color}
\usepackage[T1]{fontenc}

\usepackage[normalem]{ulem}

\newcommand{\e}{\ensuremath{\mathrm{e}}}
\newcommand{\di}{\ensuremath{\mathrm{d}}}
\newcommand{\R}{\ensuremath{\mathbb{R}}}
\newcommand{\us}[1]{\ensuremath{\frac{1}{#1}}} 

\title{Revealing intermittency in experimental data with steep power spectra}
\author{E. Falcon\inst{1} \and S. G. Roux\inst{2} \and B. Audit\inst{3}}
\shortauthor{E. Falcon \etal}

\institute{
  \inst{1} Laboratoire Mati\`ere et Syst\`emes Complexes (MSC), Universit\'e Paris Diderot, CNRS -- UMR 7057\\ 10 rue A. Domon \& L. Duquet, 75013 Paris, France, EU\\
  \inst{2} Universit\'e de Lyon, Laboratoire de Physique, CNRS, Ecole Normale Sup\'erieure de Lyon, 69007 Lyon, France, EU\\
  \inst{3} Universit\'e de Lyon, Laboratoire Joliot-Curie, CNRS, Ecole Normale Sup\'erieure de Lyon, 69007 Lyon, France, EU
}

\abstract{
The statistics of signal increments are commonly used in order to test for
possible intermittent properties in experimental or synthetic data.
However, for signals with steep power spectra 
[i.e., $E(\omega) \sim \omega^{-n}$ with $n \geq 3$], the increments are poorly
informative and the classical phenomenological relationship between
the scaling exponents of the second-order structure function and of the
power spectrum does not hold.
We show  that in these conditions the relevant quantities to compute are the second or higher degree differences of the signal.
Using this statistical framework to analyze a synthetic signal and
experimental data of wave turbulence on a fluid surface, we accurately
characterize intermittency of these data with steep power spectra.
The general application of this methodology to study
intermittency of experimental signals with steep power spectra is
discussed.}

\begin{document}
\maketitle

\section{Introduction}
Since the prediction of Kolmogorov in 1941 \cite{Kolmogorov41}, it is
well known that the spatial power spectrum $E(k)$ of a fluid particle
velocity $v$ in a turbulent flow is a power-law of the wave number $k$
as $k^{-5/3}$. The $-5/3$ exponent of the spatial power spectrum is
related to the second-order moment of velocity increments $S_2(r)
\equiv \langle [v(l+r)-v(l)]^2 \rangle \sim r^{2/3}$, $l$ and $r$
being a position and a spatial separation~\cite{Pope}. The
phenomenological relationship between both exponents comes from
Fourier transform properties. It can be generalized to any stationary
random processes: if the power spectrum of the process is $E(k)\sim
k^{-n}$, then $S_2(r) \sim r^{\zeta_2}$ with  $\zeta_2=n-1$~\cite{Pope}.
This property allows to perform measurements in the real
space to reach the power-law exponent of the power spectrum. 
The statistics of velocity increments are also crucial to characterize the intermittent nature of the velocity
field using the scaling properties of structure functions:
$S_p(r) \equiv \langle [v(l+r)-v(l)]^p \rangle \sim r^{\zeta_p}$ ($p$
positive integer)~\cite{ParisiFrisch1985}.
A non-linear dependence of $\zeta_p$ versus $p$ is the hallmark
of intermittency.

Steep power-law spectra ($\sim \omega^{-n}$ or $\sim k^{-n}$ with $n$
close or larger than 3) of a process are observed in various
situations: magnetohydrodynamics turbulence~\cite{Petrelis},
atmospherics turbulence~\cite{Julian70}, gravity~\cite{Toba73} or
capillary~\cite{Falcon07} wave turbulence on a fluid surface, and direct
cascade of two-dimensional fluid turbulence~\cite{Paret99,Rutgers98}.
Whatever the corresponding signal measured in space or in time ({\it e.g.}
fluid velocity or vorticity, surface-wave height, magnetic field, wind),
such steep spectra mean that the measured signal is at least once continuously
differentiable~\cite{Pope}. The signal differences or increments are thus
poorly informative since they are dominated by the differentiable
component of the signal.
For instance, some numerical simulations of the power-law scaling
of the energy spectrum in two-dimensional turbulence exhibited apparent contradictions with its reconstruction from spatial correlation
measurement, i.e. $\zeta_2 \neq n-1$ (see Ref. \cite{Babiano85}).
Babiano~\emph{et~al.} have  systematically
reconsidered the theoretical relations between second-order structure
functions and energy spectra instead of phenomenological or
dimensional arguments~\cite{Babiano85}. They showed that the apparent
contradictions come from the fact that the relation $\zeta_2=n-1$ does
not hold for steep power-law spectra.
Indeed, due to the differentiable component, the exponent of the
second-order structure function is independent of the spectrum slope
as soon as this one is steeper than $-3$; that is $\zeta_2=2$ whatever
$n\geq3$~\cite{Babiano85}. This latter property has been noted elsewhere without derivation~\cite{Pope,MoninYaglom,Davidson05}. The indirect conclusions drawn from the structure function analysis to reach the exponent of the power-law spectrum (using the relation
$\zeta_2=n-1$) are thus misleading for $n\geq3$, but suprinsingly are still used for some experimental signals with steep power spectra~\cite{Kellay95}. 

In this letter, we emphasize that the increments are not the relevant
quantities in order to statistically characterize the fluctuations of a steep power spectrum signal and, in particular, to probe for its possible intermittent nature.
We show analytically that, depending on the spectral steepness, it is
necessary to adapt the degree of the difference statistics used to analyze 
the signal.
Given an adapted degree, we provide a general relationship between the
exponents of the second-order structure function and of the power
spectrum whatever the steepness of the spectrum. Finally, applying this
approach to a synthetic signal, and to an experimental signal of
wave turbulence on a fluid surface allows us to accurately characterize intermittency
of these data with steep power spectra. Note that a general framework for the study of intermittency of a signal with
arbitrary degree of regularity has been previously proposed using more
complex estimators based on the continuous wavelet transform~\cite{WTMM}
or on inverse statistics~\cite{Jensen99}.
Here, we propose a practical approach and provide simple rules that should
be easily applicable on the workbench to study intermittency of
experimental signals with steep spectra. 

\section{Scaling properties of irregular signals}
Let us first recall the pioneering work of Parisi and Frisch~\cite{ParisiFrisch1985}
introduced for the description of the irregular nature of longitudinal velocity data in
fully developed turbulence.
They locally described the fluctuations of an erratic signal, $\eta(t)$, by means of
the singularity exponents $h(t_0)$ which caracterizes the power-law behaviour
of the finite differences (or increments) of $\eta$ over a time lag $\tau$ at a time $t_0$
\begin{equation}
  \label{def:sing}
  \delta_{\tau} \eta(t_0) \equiv \eta(t_0+\tau)-\eta(t_0)
  \underset{\tau\mapsto0^+}{\propto}
  \tau^{h(t_0)}\;.
\end{equation}
Since the characterization of non-local singularities cannot be achieved
in a purely local manner, Parisi and Frisch introduced the $p$-order structure 
functions
\begin{equation}
  \label{def:sf}
  S_p(\tau)\equiv \left\langle |\delta_{\tau}\eta(t)|^p \right\rangle\;,
\end{equation}
where $\langle \cdot \rangle$ represents an average over time $t$ and
from which they defined the spectra of global exponents $\zeta_p$ such as
\begin{equation}
  \label{def:zp}
  S_p(\tau)\propto \tau^{\zeta_p}\;.
\end{equation}
Note that it is useful to consider the moment of order $p$ of $|\delta_{\tau}\eta(t)|$
rather than $\delta_{\tau}\eta(t)$ as used in Ref.~\cite{ParisiFrisch1985}, so that Eqs.~(\ref{def:sf}) and~(\ref{def:zp})
are defined for all $p$ positive real.
In the presence of homogeneous fluctuations, i.e.\ $h(t_0)=H$ whatever $t_0$, it is
straightforward that $\zeta_p$ is a linear function of $p$ with $\zeta_p=pH$.
Reciprocally, a non-linear dependence of $\zeta_p$ with $p$ is the signature
of non-homogeneous fluctuations whose singularity exponents vary with time, i.e. the hallmark of intermittency.

\section{Scaling exponents of two point statistics}
We now resume the relationships between the power-law
exponents of the second-order structure
functions, the correlation function, and the power spectrum.
Let us consider a stochastics process $\eta(t)$ which is not necessarily
stationary (see below).
We assume that finite differences of $\eta$ over a time lag $\tau$,
$\delta_{\tau} \eta(t) \equiv \eta(t+\tau)-\eta(t)$, form a stationary process of
zero mean and that $\eta(0)=0$.
Let us look at the correlation between two short intervals (of size $\Delta$)
separated by a lag $\theta$, that is the quantity
$\delta_{\Delta}\eta(t+\theta)$$\delta_{\Delta}\eta(t)$.
We choose $\Delta$ equals to the sampling time of $\eta(t)$,  that is
the duration  between two successive measurement points.
For $p=2$, the scaling of Eq.~(\ref{def:zp}) implies a power-law decay
of the correlation function  $C$ of the increments $\delta_{\Delta}\eta$
at large lags $\theta$ as \cite{SamorodniskyTaqqu1994}
\begin{equation}
  \label{def:cor}
  C(\theta)\equiv \left\langle \delta_{\Delta}\eta(t+\theta) \delta_{\Delta}\eta(t) \right\rangle
  \underset{|\theta|\mapsto\infty}{\propto}|\theta|^{-\kappa}\;,
  \text{with } \kappa=2-\zeta_2\;.
\end{equation}
This scaling behaviour for large lags coincides to a low-frequency power-law
behaviour of the power spectrum $\widehat{C}$ of the increments $\delta_{\Delta}\eta$,
defined as the Fourier transform (FT) of the correlation 
function \cite{SamorodniskyTaqqu1994}
\begin{equation}
  \label{def:cspec}
  \widehat{C}(\omega)\equiv\int_\R C(\theta)\e^{i\omega\theta}\di\theta
  \underset{|\omega|\mapsto0}{\propto}|\omega|^{-\beta}\;,
  \ \text{with }\ \beta=\zeta_2-1\;,
\end{equation}
$\omega$ being the angular frequency.
An estimator of the power spectrum $\widehat{C}(\omega)$ is obtained taking the
square modulus of the FT of the increments $\delta_{\Delta}\eta(t)$ observed
over a finite time range $[0,T]$,
\begin{equation}
  \label{eq:spectrum}
  \widehat{C}(\omega)\simeq E_{\delta_{\Delta}\eta}(\omega)\equiv \left|\us{T}\int_0^T \delta_{\Delta}\eta(t)\e^{i\omega t}\di t\right|^2\;.
\end{equation}
In practice, a common habit is to compute the empirical power spectrum of $\eta$,
denoted $E(\omega)$, rather than that of the increments, $E_{\delta_{\Delta}\eta}(\omega)$.
Indeed, these spectra are related by $E_{\delta_{\Delta}\eta}(\omega)=2\left[1-\cos(\omega)\right]E(\omega)$
leading to the empirical power spectrum scaling
\begin{equation}
  \label{def:espec}
  E(\omega)\simeq \frac{\widehat{C}(\omega)}{2\left[1-\cos(\omega)\right]}
  \underset{|\omega|\mapsto0}{\propto}|\omega|^{-n}\;,
  \ \text{with }n=\beta+2\;.
\end{equation}
The scaling exponents of $S_2(\tau)\propto \tau^{\zeta_2}$, $C(\theta)\propto |\theta|^{-\kappa}$ and $\widehat{C}(\omega)\propto \omega^{-\beta}$ for the increments thus read respectively \begin{equation}
  \label{expo:rela}
  \zeta_2=n-1{\rm ,}
  \ \ \kappa=3-n{\rm ,}
  \ {\rm and}\ \ \beta=n-2{\rm ,}
\end{equation}
where $n<3$ (see below) is the exponent of the power spectrum of $\eta(t)$ [see Eq.~(\ref{def:espec})].

\begin{largetable}
 \caption{Relationships between the spectral exponent $n$, the structure function exponent and the differentiability of a signal $\eta(t)$ with a self-similar power spectrum. Reference indicates the existing derivation of the relationships. $S^{(1)}_2(\tau)\equiv \langle|\delta_{\tau}^{(1)}\eta|^2\rangle$, $S^{(2)}_2(\tau)\equiv \langle|\delta_{\tau}^{(2)}\eta|^2\rangle$ and   $S^{(3)}_2(\tau)\equiv \langle|\delta_{\tau}^{(3)}\eta|^2\rangle$.}
   \label{tab01}
  \begin{tabular}{rlll}
    Power spectrum               & Differen- & Difference statistics used to test intermittency                                      & Second-order    \\
    $E(\omega) \sim \omega^{-n}$ & tiability &                                                                                    & structure funct. \\
    \hline
    $n < 3$               & 0         & $\delta_{\tau}^{(1)}\eta=\eta(t+\tau)-\eta(t)$                               & $S^{(1)}_2 \sim \tau^{n-1}$ \cite{Babiano85}\\
    $ n \geq 3$                  & $\geq 1$  & -                                                                            & $S^{(1)}_2 \sim \tau^{2}$ \cite{Babiano85}  \\
    \hline
    $n < 5$               & 1         & $\delta_{\tau}^{(2)}\eta=\eta(t+2\tau)-2\eta(t+\tau)+\eta(t)$                & $S^{(2)}_2 \sim \tau^{n-1}$  \\
    $ n \geq 5$                  & $\geq 2$  & -                                                                            & $S^{(2)}_2 \sim \tau^{4}$  \\
    \hline
    $n < 7$               & 2         & $\delta_{\tau}^{(3)}\eta=\eta(t+3\tau)-3\eta(t+2\tau)+3\eta(t+\tau)-\eta(t)$ & $S^{(3)}_2 \sim \tau^{n-1}$ \\
    $ n \geq 7$                  & $\geq 3$  & -                                                                            & $S^{(3)}_2 \sim \tau^{6}$ 
  \end{tabular}
\vspace*{-.8cm}
  \end{largetable}

It is fundamental to note that the above stationarity condition for the increments
does not imply the stationarity of the signal, so that the correlation function
of $\eta$ might not be defined and, thus, that $E$ may not be a power spectrum
in the statistical sense ({\it i.e.} the FT of a correlation function).
Also, one should be careful that the estimation of the spectrum $E(\omega)$
for a non-stationary signal $\eta(t)$ may be significantly biased depending on
the FT numerical algorithm used. In such condition, a more robust practice to
estimate the spectral scaling exponent $n$ consists in numerically computing
the FT of the stationary signal $\delta_{\Delta}\eta(t)$, {\it i.e.},
$E_{\delta_{\Delta}\eta}(\omega)$, and to use Eq.~(\ref{def:espec}) to convert
back to the usual power spectrum.

\section{Meaning of steep power spectra}
Here, we explain why the usual relationships of Eq.~(\ref{expo:rela}) no longer hold
in the presence of steep power spectra ($n\geq3$).
In this case, Eq.~(\ref{expo:rela}) leads to $\kappa<0$, suggesting that the correlation
function of $\delta_{\Delta}\eta$ does not go to 0 but rather diverge at large lag
values [see Eq.~(\ref{def:cor})].
Obviously, this is neither physically nor statistically acceptable.
Indeed, for a non trivial stationary process $\delta_{\tau}\eta$ satisfying
Eq.~(\ref{def:zp}), the correlation function of Eq.~(\ref{def:cor}) is only defined for
$\kappa=2-\zeta_2>0$, that is for $\zeta_2 < 2$.
This implies that the spectrum of the increments is only defined  for $\beta=\zeta_2-1<1$.
Consequently, for a process with stationary increments, the scaling exponent of the
empirical power spectrum must satisfies $n=\zeta_2+1<3$.
Hence, for steep power spectra ($n\geq3$), the basic assumption that increments form
a stationary process is not verified so that the structure functions of
Eq.~(\ref{def:zp}) and the correlation function of Eq.~(\ref{def:cor}) are not well
defined.
In practice, the classical phenomenological relation between
$E(\omega)\sim \omega^{-n}$ and $S_2(\tau) \sim \tau^{n-1}$ is thus invalid for
$n \geq 3$, so that the spectral slope can not be deduced from the measurement of the
second-order structure function.
This also means that the process $\eta(t)$ is at least once differentiable
at times where
$\eta(t+\tau)-\eta(t)\simeq \tau d\eta/dt$ at the first order in $\tau$ \cite{Pope}.
These local linear trends are responsible for the non-stationarity of
signal increments and, in turns, bias the estimation of scaling exponents.
Indeed, near these times, one has $|\delta_{\Delta}\eta(t)|^p \sim \tau^p$ that corresponds to $\zeta_p = p$ [using Eqs.~(\ref{def:sf}) and (\ref{def:zp})]. This means that the scaling of the exponent of the structure functions are independent of the spectral steepness. Thus, the increments of the signal do not appear as relevant quantities when looking for possible intermittency (i.e. a non-linear evolution of $\zeta_p$ with $p$) in signals with steep power spectra. 

\section{Using higher-degree difference statistics to recover stationarity}
As recalled above, the scaling exponent of the spectrum of the increments is decreased by
two with respect to the one of the spectrum of the signal ($\beta=n-2$).
Clearly, the repetition of the difference process allows recovering the power spectrum of
a stationary process.
For instance, for $3\leq n<5$, the second-degree difference of the signal
$\delta_{\Delta}^{(2)}\eta(t)\equiv \delta_{\Delta}[\delta_{\Delta}\eta(t)]=\eta(t+2\Delta)-2\eta(t+\Delta)+\eta(t)$
has a power spectrum with scaling exponent $\beta^{(2)}=n-4<1$, which is compatible with
$\delta_{\Delta}^{(2)}\eta(t)$ being stationary.
The second-degree differences thus remove the local linear trends in the signal $\eta$ responsible for
the saturation $\zeta_p = p$ when $n\geq3$. Thus, when looking for possible intermittency in this case, one should use the structure functions of degree 2, that is
$S^{(2)}_p(\tau)\equiv \langle|\delta_{\tau}^{(2)}\eta(t)|^p\rangle\propto|\tau|^{\zeta^{(2)}_p}$.
We have indeed  $\zeta^{(2)}_2 = n-1$ for $n<5$. For $n\geq5$, one have $\zeta^{(2)}_2 = 4$  due to local quadratic trends in the signal. Thus, for $5\leq n <7$, the power spectrum of third-degree differences 
$\delta_{\Delta}^{(3)}\eta(t)\equiv\delta_{\Delta}[\delta^{(2)}_{\Delta}\eta(t)]=\eta(t+3\Delta)-3\eta(t+2\Delta)+3\eta(t+\Delta)-\eta(t)$,
is well defined and intermittency should be tested with 
$S^{(3)}_p(\tau)\equiv \langle|\delta_{\tau}^{(3)}\eta(t)|^p\rangle\propto|\tau|^{\zeta^{(3)}_p}$.
These results are summarized in the Table~\ref{tab01}.
Note that, even though the slope of the power spectrum is reduced by computing
higher-degree difference statistics, the upper bound of the spectral bandwidth, related
to the finite dynamic resolution of the original signal measurement, is not bypassed.

\section{Workbench recipe}
In practice, given a first estimate of the empirical power spectrum scaling exponent $n$,
further statistical analysis should be performed  using difference statistics of degree $d>d^*$
where $d^*$ is the smallest integer such that $n-2d^*<1$. For instance, structure function of degree $d$,
$S^{(d)}_p(\tau)\equiv \langle|\delta_{\tau}^{(d)}\eta(t)|^p\rangle\propto|\tau|^{\zeta^{(d)}_p}$
should be used to test for intermittency while spectral analysis or
correlation analysis should be done
on degree $d$ differences, $E_{\delta_{\Delta}^{(d)}\eta}(\omega)\propto|\omega|^{\beta^{(d)}}$
and $C_{\delta_{\Delta}^{(d)}\eta}(\theta)\propto|\theta|^{-\kappa^{(d)}}$.
The relationships between the scaling exponents then become 
\begin{equation}
  \label{expo:relabis}
  \zeta_2^{(d)}=n-1{\rm ,}
  \ \ \kappa^{(d)}=1+2d-n{\rm ,}
  \ \ \beta^{(d)}=n-2d{\rm ,}
\end{equation}
for $n < 1+2d$. In other words, the proposed procedure removes the biases in the estimation of the scaling exponents. These biases arise from the local regular behaviors of a signal that can occur at different degrees of the signal differentiability. Hence, in order to numerically check that the signal regular components have been adequately removed, it is good practice to check
that results remains consistent when increasing degree $d$ to $d+1$.
In particular, one should make sure that $\zeta_p^{(d)}\simeq \zeta_p^{(d+1)}$. Note that for discontinuous signals the use of increments or higher-degree differences is unsuitable, a wavelet-based approach is more suited \cite{WTMM}.

\section{Applications}
To illustrate the results of the previous section, let us now apply the
proposed estimator based on higher-degree difference statistics to signals
with steep spectra and probe their possible intermittent nature.
A synthetic signal with prescribed intermittency and experimental data
of wave turbulence on a fluid surface will be tested below for comparison.
To our knowledge, only one study has compared the method of second-degree differences with more complex estimators based on inverse statistics in order to probe intermittency in a simulation of two-dimensional flows~\cite{Biferale03}.

\begin{figure}[t]
\onefigure[width=\linewidth]{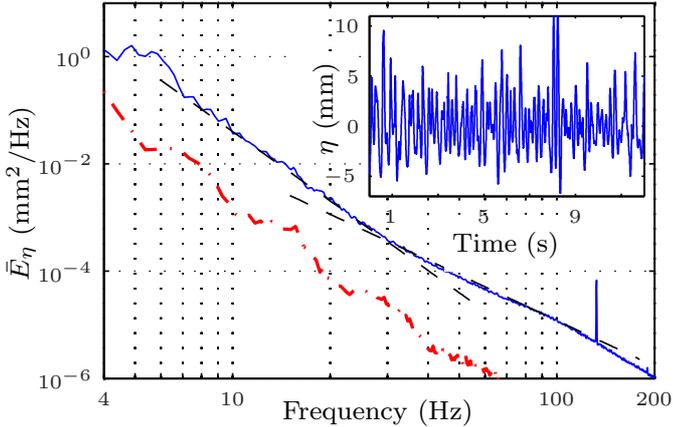}
\caption{Power spectra of the experimental data $\eta(t)$ (solid line)
  and of a synthetic signal (dash-dotted line). Dashed lines have slopes
  of -4.3 and -2.8. Inset: Typical temporal evolution of $\eta(t)$ during
  10~s, $\langle \eta \rangle \simeq 0$.}
\label{Fig01}
\end{figure}

\begin{figure}[t]
\onefigure[width=\linewidth]{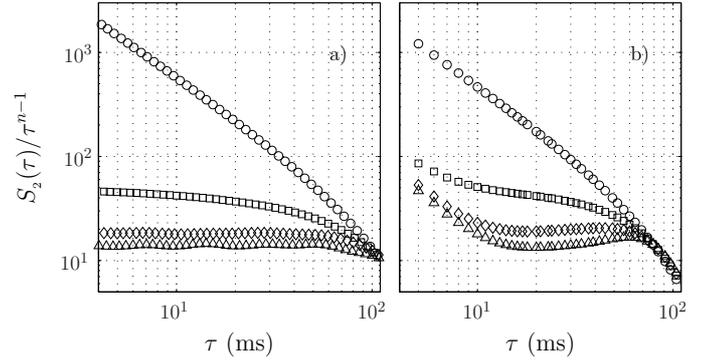}
\caption{Rescaled second-order moment of the structure functions ${\mathcal S}_2(\tau)/\tau^{n-1}$ with $n=4.3$ computed from the ($\circ$) first-, ($\square$) second-, ($\diamond$) third- and ($\triangle$) fourth-degree differences of the signal as a function of the time lag $\tau$. a)  Synthetic signal. b) Experimental signal. 
Correlation time is $\tau_c\simeq 63$ ms.}
\label{Fig02}
\vspace*{-.3cm}
\end{figure}

\begin{figure}[t]
\onefigure[width=\linewidth]{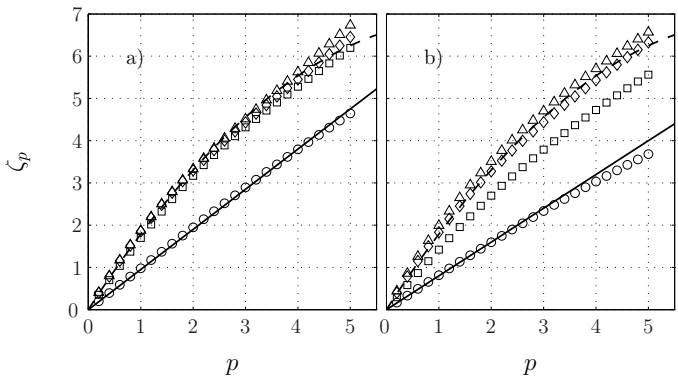}
\caption{Structure function exponents, $\zeta_p$, as a function of $p$ for the synthetic signal (a) and experimental data (b). ($\circ$) $\zeta_p^{(1)}$ computed from the first-degree increments ; ($\square$) $\zeta_p^{(2)}$ computed from the second-degree differences and fitted by ($--$) $\zeta_p^{(2)}=c_1p-c_2p^2/2$ with $c_1=1.92$ and $c_2=0.27 $; 
($\diamond$) $\zeta_p^{(3)}$ and ($\triangle$) $\zeta_p^{(4)}$ computed from the third-degree and fourth-degree differences. 
Solid lines are linear fits of $\zeta_p^{(1)}$: (a) $\zeta_p^{(1)}=0.95p$, and (b) $\zeta_p^{(1)}=0.8p$.}
\label{Fig03}
\vspace*{-.3cm}
\end{figure}

\subsection{Synthetic data}
Here, we apply the above suggested estimator to a synthetic data. Recently, it has been proposed that the scaling properties of experimental
velocity (transverse to the mean flow) in fully developed turbulence could
be described by log-normal Random Wavelet Cascade (RWC)~\cite{amr97}. 
RWC generalizes the concept of self-similar cascades leading to multifractal measures  ($-1 \leq n \leq 1$) to the construction of scale-invariant signals ($n>1$) using orthonormal wavelet basis~\cite{amr97}. Instead of redistributing the measure over sub-intervals with multiplicative weights, it allocates the wavelet coefficients in a multiplicative way on the dyadic grid. This method has been implemented to generate multifractal functions  from a given deterministic or probalistic multiplicative process. From a mathematical point of view, the convergence of the cascade and the regularity properties of the so-obtained stochastic functions have been discussed in Ref.~\cite{abm98}. 
Intermittency of RWC is characterized by the theoretical scaling exponents
$\zeta_p=c_1p-c_2p^2/2$~\cite{abm98}. 
Here, we consider a realization of $3\times 10^6$ data points of the RWC process and choose $c_1=1.92$ and $c_2=0.27$ 
to reproduce the  intermittent properties of experimental data of wave turbulence (see below). 

\begin{figure}
    \onefigure[width=0.99\linewidth]{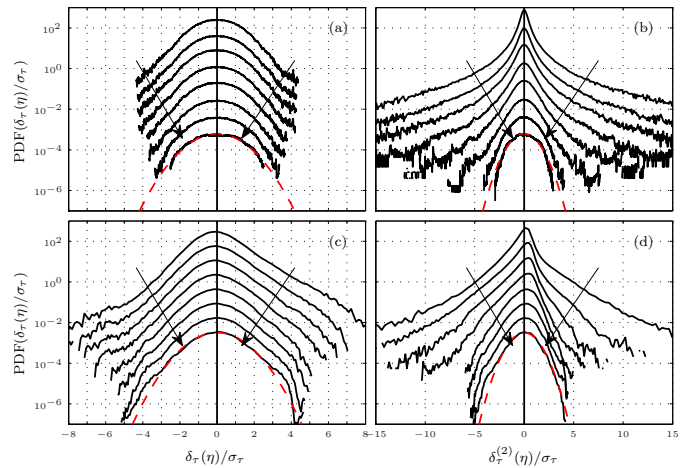}
  \caption{PDFs of the first-degree increments $\delta_{\tau}^{(1)}\eta$ for different $\tau$ from 6 to 100 ms (see arrows): (a) Synthetic data, and (c) Experimental data. PDFs of the second-degree differences $\delta_{\tau}^{(2)}\eta$ for $6 \leq \tau \leq 100$ ms: (b) Synthetic data, and (d) Experimental data. Dashed-line: Gaussian with zero mean and unit standard deviation. $\sigma_{\tau}$ are the rms values. Each curve has been shifted vertically for clarity.}
  \label{Fig04}
\vspace*{-.3cm}
\end{figure}

We first compute the usual power spectrum of the RWC signals allowing us to assess
that its frequency power law exponent is around $3<n<5$.
This suggests that unbiased estimates of the scaling exponents will be obtained using
second (or higher) degree difference statistics (see Table~\ref{tab01}).
The unbiased power spectrum of this signal is then computed using the Fourier transform [see Eqs.~(\ref{eq:spectrum}) and~(\ref{def:espec})] of the second-degree differences  and is shown in Fig.~\ref{Fig01}.
It roughly behaves as a steep power-law over one decade in frequency,
i.e. $E(\omega) \propto \omega^{-n}$ with $n \simeq 4.3$. 
The second-order structure functions $S_2^{(d)}$ of this synthetic signal are then computed using the first, second, third and fourth degree difference statistics ($d=1$, 2, 3  and 4) as shown in Fig.~\ref{Fig02}a.
We observe that, when using the third-degree statistics ($d=3$), one has $S_2^{(3)}(\tau)\simeq\tau^{\zeta_2^{(3)}}$ with $\zeta_2^{(3)}=3.33$ that is in good agreement with the
theoretical value $\zeta_2=2(c_1-c_2)=3.3$, the classical
relationship $\zeta_2^{(3)} = n-1$ being thus well satisfied.
When $d=4$, we obtain $\zeta_2^{(4)}=3.35$ a value that is consistent with the one found with $d=3$.
For $d=2$, one obtains $\zeta_2^{(2)}=3.23$, a value that is slightly below the previous one since it begins to be biased by the smoother part of the signal (two times differentiable or more). Finally, using the usual first-degree statistics ($d=1$) leads to
$\zeta_2^{(1)}\simeq2$ since the RWC signal is differentiable with a steep power spectrum of exponent $n\geq3$ (see Table~\ref{tab01}). 

The estimations of the structure-function exponents $\zeta_p^{(d)}$ versus $p$ using the first, second, third and fourth degree difference statistics ($d=1$, 2, 3  and 4) are presented in Fig.~\ref{Fig03}a.
We observe that $\zeta_p^{(2)}$ is a non-linear function of $p$ which
provides a clear evidence of the intermittent nature of the RWC signal
fluctuations.
The fact that $\zeta_p^{(3)}$ and $\zeta_p^{(4)}$ estimates are consistent with the previous ones ($\zeta_p^{(2)}\simeq \zeta_p^{(3)}\simeq \zeta_p^{(4)}$) provides further confidence on this
diagnosis.
It is noteworthy that $\zeta_p^{(2)}$, $\zeta_p^{(3)}$ and $\zeta_p^{(4)}$ estimates are in
good agreement with the theoretical expectation $\zeta_p=c_1p-c_2p^2/2$,
which illustrates that the proposed framework allows an accurate
characterization of the intermittent properties.
Finally, as expected for a differentiable signal, using
first-degree increments leads to $\zeta_p^{(1)}$ is a linear function of $p$
with a slope close to 1. This latter result thus leads to a misleading conclusion that the RWC signal does not present intermittency. This exemplifies the need to adapt the degree of the difference statistics to the steepness of the power spectrum.

Finally, the probability density functions (PDFs) of the first-degree increments of
the synthetic signal are plotted in Fig.~\ref{Fig04}a for different time lags $\tau$.
All the PDFs have the same shape independent of the scale $\tau$, thus showing no
intermittency. In contrast, the PDFs of the second-degree increments displayed in Fig.~\ref{Fig04}b show a clear evolution across scales, highlighting the intermittency of the signal.
This is consistent with the previous structure function analysis, and 
further underlines that high-degree difference statistics is needed to test
intermittency of steep power-spectrum signals.

\subsection{Wave turbulence data}
We now apply the above proposed statistical estimator on an experimental signal
of hydrodynamics surface wave turbulence~\cite{Falcon07bis}.
A typical signal is the temporal evolution of the surface wave amplitude,
$\eta(t)$, measured at a given location of the free surface of the fluid
(see inset of Fig.~\ref{Fig01}). Data are recorded from 10 successive experiments of 300~s each where surface waves are generated by a wave maker driven by random noise forcing in a frequency range 0--6~Hz~\cite{Falcon07bis}.
Wave heights was measured at 1~kHz sampling rate ($\Delta=1$~ms)
resulting in $3\times 10^6$ data points. 
As for the RWC signal, the initial estimation of the power spectrum
steepness indicates that $3<n<5$.
In order to probe for possible intermitent properties of this signal with such a
steep power spectrum, we thus need to use the adapted difference statistics
proposed in the previous section. 

The power spectrum of $\eta(t)$, estimated using the Fourier transform [see Eqs.~(\ref{eq:spectrum}) and~(\ref{def:espec})] of $\delta_\Delta^{(2)}\eta$,  is shown in Fig.~\ref{Fig01}.
It displays two frequency ranges with a power-law behaviour.
In the low-frequency spectrum range ($\sim 7 - 30$~Hz) corresponding to the gravity
wave turbulence regime, we observe $E(\omega)\propto\omega^{-4.3}$  while the 
high-frequency range ($\sim 30-100$~Hz) corresponding to the capillary regime
is characterized by $E(\omega)\propto\omega^{-2.8}$~\cite{Falcon07}.
The second-order structure functions $S_2^{(d)}(\tau)$ of $\eta(t)$ for
$d=1$, 2, 3 and 4 are shown in Fig.~\ref{Fig02}b as a function of the time
lag $5\leq\tau\leq100$~ms.
When considering structure functions $S_2^{(d)}$ for $d\geq2$ in the time lag range
$\tau\lesssim80$~ms (corresponding to frequencies above the maximal
forcing frequency 6~Hz), we roughly observe two different power-law scaling
behaviours in the gravity regime $15 \lesssim\tau\lesssim65$ ms
and the capillary regime $5\lesssim\tau\lesssim15$~ms
[a time lag $\tau$ corresponds to a frequency $f=1/(2\tau)$].
We will focus below only to the gravity regime since the transition between both regimes in Fig.~\ref{Fig02}b occurs rather smoothly which significantly reduce the time lag range available to fit the scaling exponent in the capillary regime.
As explained above, for $d=1$, $S_2^{(1)}$ is dominated by the signal differentiability and does not display both scaling regimes. Focusing only on the gravity regime ($15 \lesssim\tau\lesssim65$ms), one consistently finds $\zeta_2^{(3)}=3.3$ and $\zeta_2^{(4)}=3.4$ in good agreement with the spectral exponent $n=4.3$. When using $d=2$, the smoother transition observed between the scaling regimes leads to a slightly underestimated value $\zeta_2^{(2)}=2.9\neq n-1$.

We then look for possible intermittent properties of the turbulence wave data in the gravity regime. The evolution of $\zeta_p^{(d)}$ with $p$ is shown in Fig.~\ref{Fig03}b for $d=1$, 2, 3 and 4. When using the first-degree increments, $\zeta_p^{(1)}\simeq0.8p$ is a linear
function of $p$.
As underlined above, in presence of a steep power spectrum, $\zeta_p^{(1)}$ is 
dominated by the differential component of the signal masking possible intermittency.
For $d=2$, 3 and 4, we observe a clear non linear behaviour of $\zeta_p^{(d)}$
versus $p$.
We note that while $\zeta_p^{(3)}$ and $\zeta_p^{(4)}$ provide consistent estimates
of $\zeta_p$ for all $p$, $\zeta_p^{(2)}$ estimates are slightly below these latters 
as already observed above for $p=2$.
Finally, the coherence between the estimates of $\zeta_p$ for two successive values
of the difference degree ($\zeta_p^{(3)}\simeq \zeta_p^{(4)}$) and with the spectral
analysis ($\zeta_2^{(3,\; 4)}\simeq n-1$) strongly suggests that these measurements are reliable.
We can thus conclude these data of wave turbulence are intermittent. Fitting of $\zeta_p^{(3)}$ with the polynomial model $\zeta_p=c_1p-c_2p^2/2$ yields $c_1=1.9$ and an intermittency coefficient $c_2=0.27$.

Another way to highlight the wave turbulence intermittency is to observe a shape
deformation of the probability density functions (PDFs) of the signal increments
with the time lag $\tau$.
The PDFs of the first and second-degree increments of the wave amplitudes are
respectively plotted in Figs.~\ref{Fig04}c and \ref{Fig04}d for
$6 \leq \tau \leq 100$ ms.
When $\tau$ is increased, the PDF's shape of the second-degree increments changes
continuously up to a nearly Gaussian shape at large $\tau$ (see Fig.~\ref{Fig04}d).
This deformation is a direct signature of intermittency.
As predicted above, this intermittency is not diagnose when using the
first-degree estimator: almost no deformation of the PDF shapes of the first-degree increments
is observed in Fig.~\ref{Fig04}c. 
 
In this Letter, we have proposed an easily applicable framework based on
high-degree difference statistics to probe for possible intermittency of a signal with a steep power spectrum.
We applied it to synthetic data and to wave turbulence data.
This has led to the observation of wave turbulence 
intermittency~\cite{Falcon07bis}.
In the same way, it can be used on previous existing data notably of
magnetohydrodynamic~\cite{Petrelis} or two-dimensional turbulence~\cite{Paret99}, both showing steep power spectra.

\acknowledgments
We thank A. Newell, F. P\'etr\'elis, and A. Arneodo for fruitfull discussions. This work has been supported by ANR Turbonde BLAN07-3-198746.



\begin{thebibliography}{99}
  
\bibitem{Kolmogorov41}
  \Name{Kolmogorov A. N.}
  \REVIEW{Dokl. Akad. Nauk SSSR}{30}{1941}{9}; {\bf 32} (1941) 16, reproduced in {\it Proc. R. Soc. Lond. A} {\bf 434} (1991) 9; {\bf 434} (1991) 15
\bibitem{Pope}
  \Name{Pope S. B.}
  \Book{Turbulent Flows}
  \Publ{Cambridge University Press}
  \Year{2000}
  \bibitem{ParisiFrisch1985}
\Name{Frisch U.}
 \Book{Turbulence}
 \Publ{Cambridge University Press}
 \Year{1995}. This approach was first introcuced in
  \Name{Parisi G.\and Frisch U.}
  \Book{Proc. Intern. School on Turbulence and Predictability in Geophysical Fluid Dynamics and Climate Dynamics}
  \Editor{M. Ghil, R. Benzi, \and G. Parisi}
  \Publ{North-Holland, Amsterdam}
  \Year{1985}
  \Page{84}.

  \bibitem{Petrelis}
  \Name{Bourgoin M. et al.}
\REVIEW{Phys. Fluids}{14}{2002}{3046}.
\bibitem{Julian70}
  \Name{Julian R., Whasington W. M., Hembree, L. \and Ridley C.}
  \REVIEW{J. Atmos. Sci.}{27}{1970}{376};
  \Name{Morel P. \and Larcheveque M.}
  \REVIEW{J. Atmos. Sci.}{31}{1974}{2189};
  \Name{Desbois M.}
  \REVIEW{J. Atmos. Sci.}{32}{1975}{1838}
  \bibitem{Toba73}
  \Name{Toba Y.}
  \REVIEW{J. Oceanogr. Soc. Jpn.}{29}{1973}{209}; K. K. Kahma, {\it J. Phys. Oceanogr.} {\bf 11}, 1503 (1981); G. Z. Forristall, {\it J. Geophys. Res. Oceans Atmos.} {\bf 86}, 8075 (1981); M. A. Donelan {\it et al.}, {\it Philos. Trans. R. Soc. A} {\bf 315}, 509 (1985)
\bibitem{Falcon07}
  \Name{Falcon E., Laroche C. \and Fauve S.} 
  \REVIEW{Phys. Rev. Lett.}{98}{2007}{094503}, and references therein.
\bibitem{Paret99}
  \Name{Paret J., Jullien M.-C. \and Tabeling P.}
  \REVIEW{Phys. Rev. Lett.}{83}{1999}{3418};
  \Name{Tabeling P.}
  \REVIEW{Phys. Reports}{362}{2002}{1}.
\bibitem{Rutgers98}
  \Name{Rutgers M. A.}
  \REVIEW{Phys. Rev. Lett.}{81}{1998}{2244}, and references therein.
\bibitem{Babiano85}
  \Name{Babiano A., Basdevant C. \and Sadourny R.}
  \REVIEW{C. R. Acad. Sc. Paris}{299}{1984}{495} (in french);
  \REVIEW{J. Atmos. Sci.}{42}{1985}{941} and references therein
\bibitem{Davidson05}
  \Name{Davidson P. A. \and Pearson B. R.}
  \REVIEW{Phys. Rev. Lett.}{95}{2005}{214501}
\bibitem{MoninYaglom}
  \Name{Monin A. S. \and Yaglom A. M.}
  \Book{Statistical Fluid Mechanics: Mechanics of Turbulence}
  \Vol{2}
  \Publ{The MIT Press}
  \Year{1975}
  
\bibitem{Kellay95}
  \Name{Kellay H., Wu X-l. \and Goldburg W. I.}
  \REVIEW{Phys. Rev. Lett.}{74}{1995}{20};
  \Name{Riviera M., Vorobieff P. \and Ecke R. E.}
  \REVIEW{Phys. Rev. Lett.}{81}{1998}{1417}
  
\bibitem{WTMM}
  \Name{Muzy J. F., Bacry E. \and Arneodo A.}
  \REVIEW{Phys. Rev. Lett.}{67}{1991}{3515};
  \REVIEW{Phys. Rev. E}{47}{1993}{875};
  \REVIEW{Int. J. Bifurc. Chaos}{4}{1994}{245-302}


\bibitem{Jensen99}
  \Name{Jensen M. H.}
  \REVIEW{Phys. Rev. Lett.}{83}{1999}{76}
  
\bibitem{SamorodniskyTaqqu1994}
  \Name{Samorodnisky G. \and Taqqu M.~S}
  \Book{Stable Non-Gaussian Random Processes}
  \Publ{Chapman and Hall, New York}
  \Year{1994}.

\bibitem{Biferale03}
  \Name{Biferale L., Cencini M., Lanotte A. S. \and Vergni D.}
  \REVIEW{Phys. Fluids}{15}{2003}{1012}

\bibitem{amr97} 
  \Name{Arneodo A., Muzy J.~F. \and  Roux S.G.} 
  \REVIEW{J. Phys. II France}{7}{1997}{363}

\bibitem{abm98} 
  \Name{Arneodo A., Bacry E. \and Muzy J.~F.} 
  \REVIEW{J. Math. Phys.}{39}{1998}{4142}

\bibitem{Falcon07bis}
  \Name{Falcon E.,  Fauve S. \and Laroche C.} 
  \REVIEW{Phys. Rev. Lett.}{98}{2007}{154501}

\end{thebibliography}
\end{document}